\documentclass[APS, preprint,superscriptaddress]{revtex4-2}
\usepackage{amsmath}
\usepackage{bm}
\usepackage{physics}

\usepackage{array}
\usepackage{dcolumn}
\usepackage{adjustbox}
\newcolumntype{C}[1]{>{\centering\arraybackslash}m{#1}}

\usepackage{graphicx}
\usepackage{rotating}
\usepackage[caption=false]{subfig}
\usepackage{setspace}

\usepackage{placeins}

\begin{document}
\preprint{APS/123-QED}
\title{Simulating Exciton Transport with Complex Absorbing Potentials}
\author{Dimitri Bazile}
\affiliation{Department of Chemistry and Biochemistry, University of California, Los Angeles, Los Angeles, CA 90095, USA}
\author{Justin Caram}
\affiliation{Department of Chemistry and Biochemistry, University of California, Los Angeles, Los Angeles, CA 90095, USA}
\author{Chern Chuang}
\affiliation{Department of Chemistry and Biochemistry, University of Nevada, Las Vegas, NV 89154, USA}
\author{Daniel Neuhauser}
\email{dxn@ucla.edu}
\affiliation{Department of Chemistry and Biochemistry, University of California, Los Angeles, Los Angeles, CA 90095, USA}
\date{\today}
\begin{abstract}
We introduce a stochastic framework based on complex absorbing potentials (CAPs) to investigate exciton transport in large molecular aggregates.  Within this approach, CAPs act as non-Hermitian reservoirs and sinks that enable effective measurement of transport efficiency. We apply this framework to cyanine dye aggregates and examine how vacancy defects and system size influence exciton dynamics in two-dimensional sheets and quasi-one-dimensional tubes. We also introduce a CAPs-based classification scheme that links molecular packing in 2D aggregates to transport behavior. Our results demonstrate how aggregate topology and structural disorder govern exciton dynamics and provide guidance for designing materials with enhanced energy transport.
\end{abstract}
\maketitle
\section{Introduction}
Excitonic molecular aggregates are noncovalent assemblies of chromophores in which electronic excitations (excitons) are delocalized over multiple monomers, giving rise to photophysical properties distinct from those of isolated molecules \cite{Brixner2017}. In these systems, exciton transport arises from the interplay between coherent, wave-like propagation and incoherent hopping. This balance is strongly influenced by molecular organization, structural disorder, and packing \cite{Barford2006,Popp2021,Wong2015,Lebedenko2009}. Understanding how these factors govern exciton transport under realistic conditions is essential for improving optoelectronic materials, including organic light-emitting diodes, photovoltaic devices, and artificial photosynthetic systems. \cite{Kalinowski1997,Hedley2016,Saikin2013}.

The relative arrangement of transition dipoles determines the sign and magnitude of intermolecular couplings, giving rise to the well-known distinction between J-aggregates (head-to-tail alignment) and H-aggregates (cofacial alignment) \cite{Kasha1963}. These aggregates are typically identified by red- and blue-shifted optical spectra relative to the monomer \cite{JELLEY1937,JELLEY1936}.  More recent studies of two-dimensional (2D) and tubular cyanine dye aggregates have revealed excitonic behaviors that extend beyond this traditional classification, including I-aggregates, whose bright state is red-shifted while the band-edge structure resembles that of H-aggregates \cite{Deshmukh2022}. 

These findings highlight the importance of aggregate topology and dimensionality in determining excitonic band structure and transport properties. Although both theoretical and experimental studies suggest that exciton transport is generally more efficient in J-aggregates than in H-aggregates \cite{Wrthner2011,Petrenko2015}, spectral classification alone does not fully capture differences in transport behavior. This limitation motivates the development of transport-based classification schemes.

Energetic and structural disorder can localize excitons, a phenomenon known as Anderson localization which limits coherent transport \cite{Anderson1958,Evensky1990}. Energetic disorder arises from random variations in monomer excitation energies within an aggregate. This disorder can be spatially and temporally correlated or uncorrelated. Correlated disorder typically disrupts exciton transport less severely than uncorrelated disorder \cite{Uchiyama2018,Lee2018}. These types of disorder have been widely investigated in molecular aggregates and related materials. \cite{Akselrod2014,Reineker1993,Pant2020}.

Structural disorder can arise in several forms, including spatial variations in the relative positions and orientations of lattice sites, as well as the loss of coupling at a small number of sites due to defects \cite{DomnguezAdame1999,Biroli1999}. We refer to the latter as vacancy disorder. Vacancy effects have been investigated both theoretically and experimentally in systems such as light-harvesting nanotubes through the analysis of spectral changes induced by transient photodamage. \cite{Doria2018}. However, a comprehensive understanding of how vacancy disorder affects transport efficiency across different aggregate geometries remains unexplored.

To systematically quantify how vacancy defects and system size influence exciton transport, and to distinguish transport behavior between H- and J-aggregates, theoretical frameworks that explicitly model transport efficiency are required. In the absence of environmental coupling, exciton dynamics reduces to a unitary evolution governed by the Hamiltonian of the system \cite{Giannini2022}. For weak system–bath coupling, the dynamics are described using Redfield theory \cite{Ishizaki2009} or Lindblad master equations \cite{Mlken2010,Uchiyama2018}. However, these approaches typically exhibit high computational scaling due to the matrix operations required for time propagation, although efficient approaches have been recently developed to address these limitations \cite{Kondov2003, Adhikari2025}. 

In this work, we use complex absorbing potentials (CAPs) combined with stochastic methods to model exciton transport in molecular aggregates, enabling efficient simulations of exciton dynamics in large systems. CAPs are widely used to study metastable states with finite lifetimes, such as anions \cite{Dempwolff2021,Gyamfi2022}, to facilitate wavepacket dynamics simulations \cite{Vibok1992,Neuhasuer1989}, and to model electron transport in mesoscopic conduction systems \cite{CollepardoGuevara2004,Landauer1989,deCastro2023}. In electronic transport, CAPs can act as effective leads, allowing electrons to flow between channels and enabling the determination of transport properties in semiconductor systems \cite{Datta_1999}. More recently, CAPs have been applied to quantify exciton transport efficiency and loss rates in open quantum systems, including molecular aggregates and lossy cavity exciton transport models \cite{Sharma2025}.

In molecular aggregates, CAPs act as exciton reservoirs and sinks that collect excitons, enabling direct tracking of exciton generation, propagation, and loss. To efficiently simulate large aggregates, we employ stochastic techniques. In particular, Chebyshev expansion to approximate the action of the Green’s function \cite{Mandelshtam1995}  and stochastic trace methods are used to extract observables from the CAP formalism. These approaches circumvent full Hamiltonian operations  and have been successfully applied to compute absorption spectra and density of states (DOS) in molecular aggregate systems \cite{Bradbury2020,Bradbury2024}.

The paper is organized as follows. First, we introduce the CAP model within the Frenkel exciton framework for two-dimensional sheets and pseudo-one-dimensional tubular systems, including both one-dimensional and angle-dependent two-dimensional CAPs developed in this work. Next, we describe the stochastic methods used to compute transport metrics. In the Results section, we demonstrate that CAP-based transmission reproduces exciton delocalization trends under disorder. We then examine the effects of system size and vacancy defects in 2D sheets and tubular aggregates of TDBC and C8S3. Finally, we investigate the H–J regimes of TDBC and Cy7-DPA aggregates using angle-dependent CAPs and propose a new aggregate classification scheme. We conclude by highlighting directions for future research.

\section{Theoretical Methods}
\subsection{Frenkel Exciton Model, Vacancy and Disorder Effects}
\begin{figure}[htbp]
\centering
\includegraphics[width=0.25\linewidth]{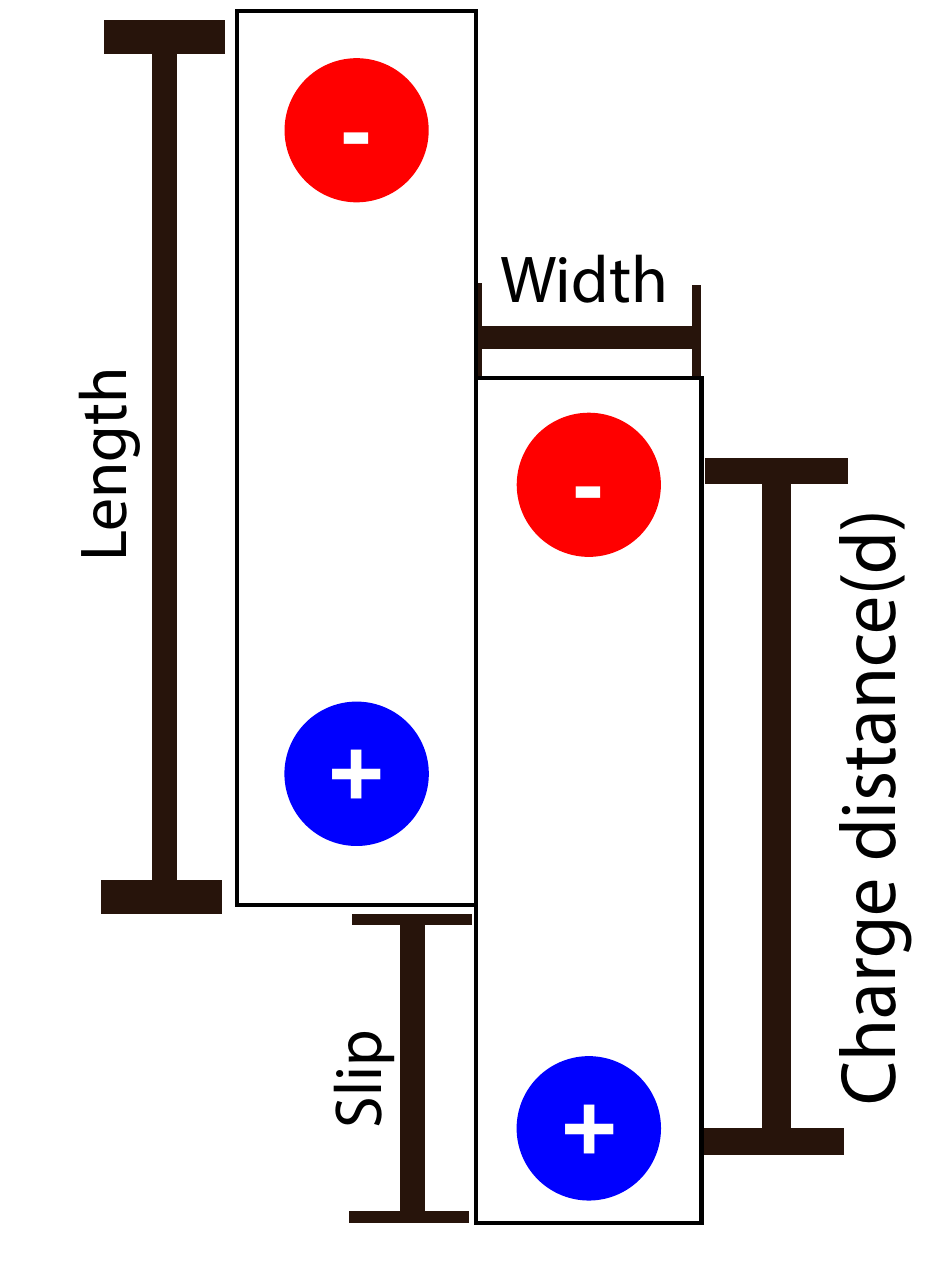}
\caption{Schematic representation of two brick-layer monomers showing the geometric parameters and extended charge distributions used to compute intermolecular coupling.}
\label{fig:brick_monomers}
\end{figure}

To model molecular aggregates with sheet and tubular structures, we employed the Frenkel exciton Hamiltonian
\begin{equation}
H_0 = \sum_n \epsilon_n |n\rangle \langle n| + \sum_{nm} J(n,m) |n\rangle \langle m|.
\label{eq:FE_MODEL}
\end{equation}
Here $J$ represents the coupling between monomers $n$ and $m$, and $\epsilon_n$ denotes the on site-energy plus the energetic disorder at monomer $n$. The coupling $J$ is calculated using the extended dipole model,
\begin{equation}
J(n,m) = \frac{1}{4D}\frac{\mu^2}{d^2}
\left(
\frac{1}{r_{m+n+}} - \frac{1}{r_{m+n-}} - \frac{1}{r_{m-n+}} + \frac{1}{r_{m-n-}}
\right).
\label{eq:coupling}
\end{equation}
$\mu$ is the effective dipole strength, $d$ is the charge separation distance, $D$ is the dielectric constant, and $r_{nm}$ is the distance between monomers $n$ and $m$. The $\pm$ subscripts denote the positions of the extended charges within each brick segment. Figure~\ref{fig:brick_monomers} illustrates these parameters for two brick monomers.

\FloatBarrier

To introduce vacancies, we modify the Hamiltonian by removing contributions from monomers at vacant sites,
\begin{equation}
H = P H_0 P .
\label{eq:vacancy}
\end{equation}
Here $P$ is a diagonal matrix of size $N_g \times N_g$, where $N_g$ is the total number of monomer sites. Each diagonal element $P_{ii}$ equals 1 if site $i$ is occupied and 0 if it contains a vacancy; all off-diagonal elements are zero. The projected Hamiltonian $H$ therefore removes all couplings to and from vacant sites. In the absence of vacancies, $P$ reduces to the identity matrix and Eq.(\ref{eq:vacancy}) recovers Eq. (\ref{eq:FE_MODEL}).

\subsection{Complex Absorbing Potentials}
Complex absorbing potentials (CAPs) are imaginary potentials added at the boundaries of a quantum system to simulate open boundary conditions. In transport calculations, CAPs mimic external leads by absorbing outgoing probability current and preventing artificial reflections at system boundaries \cite{Datta_1999,Varga2009,Xie2014}. This enables current injection at one boundary and absorption at the other. This approach is closely related to exciton transport models based on the Lindblad formalism, where one set of operators describes pumping into the excitonic manifold and another represents irreversible transfer to a sink. For example, Morales-Curiel and León-Montiel \cite{MoralesCuriel2020} studied transport efficiency in a two-dimensional J-aggregate by introducing a Lindblad sink operator and defining efficiency as the asymptotic population transferred to the sink.

\subsection{1D CAPs for Tubes and Sheets}
For two-dimensional sheet aggregates, the number of transport channels depends on the system dimensions along the $x$ and $y$ directions. The $x$  is along the brick width and the $y$ direction is along the brick length. The transport along $x$ involves $n_y$ channels (the number of monomers along $y$), while the transport along $y$ involves $n_x$ channels. Pseudo-one-dimensional tubes can be viewed as sheets rolled along the $x$ direction. In this geometry, CAPs extend along the tube height ($z$ direction) and around the circumference (rotational direction). The number of channels is $n_{\mathrm{rot}}$ along the rotational direction and $n_z$ along $z$. 
\begin{figure}[htbp]
    \centering
    \includegraphics[width=0.75\linewidth]{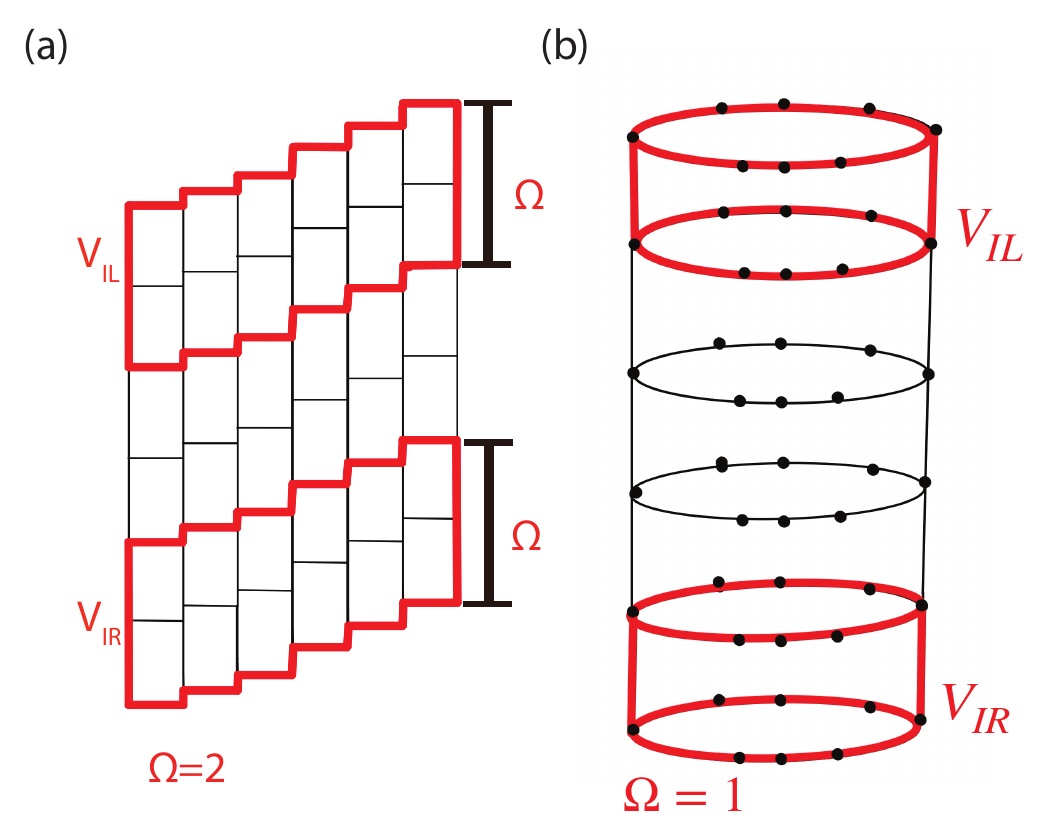}
    \caption{Schematic representation of the complex absorbing potentials (CAPs) implemented in the model systems.
(a) Two-dimensional aggregate sheet with $n_x = 6$ and $n_y = 6$. CAPs are applied along the brick width ($x$ direction) and the probability current flows along the brick length(y-direction). The monomer unit length along the y-direction $\Omega = 2$.
(b) Pseudo-one-dimensional tubular aggregate with $n_{\mathrm{rot}} = 8$ and $n_z = 6$. Black dots denote monomer centers. CAPs are applied along the rotational axis and  the probability current travels  along the tube height ($z$ direction). The monomer unit along the z-direction for CAPs  $\Omega = 1$.}
    \label{fig:caps_schematic}
\end{figure}
The CAP potentials are chosen as half-parabolas: 
\begin{equation}
\hat{V}_I= \hat{V}_{IR}+\hat{V}_{IL}    
\end{equation}
\begin{equation}
\hat{V}_{IR}= \sum_n |n  \rangle V_{IR}(w_{n}) \langle n| 
\end{equation}
\begin{equation}
\hat{V}_{IL}= \sum_n |n  \rangle V_{IL}(w_{n}) \langle n| 
\end{equation}
\begin{equation}
V_{IL}(w_n) =
\begin{cases}
V_{0I}\,\dfrac{\big(w_{\min}+d_w-w_n\big)^2}{d_w^2},
& w_{\min} \le w_n \le w_{\min}+d_w, \\[6pt]
0, & \text{otherwise}
\end{cases}
\end{equation}
\begin{equation}
V_{IR}(w_n) =
\begin{cases}
V_{0I}\,\dfrac{\big(w_n-(w_{\max}-d_w)\big)^2}{d_w^2},
& w_{\max}-d_w \le w_n \le w_{\max}, \\[6pt]
0, & \text{otherwise}
\end{cases}
\end{equation}
\( \hat{V}_I \) represents the combined left and right potentials and \( V_{OI} \) denotes the maximum height of these potentials. The chosen coordinates, $w_n$, are based on the direction of the lattice vectors.

For 2D sheet at  monomer $n$: 
\begin{equation}
\mathbf{r}_n
=
\begin{bmatrix}
x_n\\[2pt]
y_n
\end{bmatrix}
=
\begin{bmatrix}
(n_2-1)W \\[2pt]
(n_1-1)L+(n_2-1)S
\end{bmatrix}
\end{equation}
$n_1$ is the unit coordinate along the $length$ ($L$) of the brick and $n_2$ is the unit coordinate along the $width$ ($W$) and $slip$ ($S$) is the offset between adjacent monomers in neighboring rows. 

While for 1D tubes: 
\begin{equation}
\mathbf{r}_n
=
\begin{bmatrix}
x \\[2pt]
y \\[2pt]
z
\end{bmatrix}
=\begin{bmatrix}
R\cos(n_2 \phi_2+ n_1 \gamma) \\[2pt]
R\sin(n_2 \phi_2+n_1 \gamma) \\[2pt]
n_1 H 
\end{bmatrix}
\end{equation}
\begin{equation}
\phi_2=\frac{2 \pi}{n_{rot}} (n_2-1)
\end{equation}
$R$ is the radius of each circular layer, $\gamma$ is the twist angle between adjacent layers, and $H$ is the vertical spacing between layers. The indices $n_1$ and $n_2$ denote the lattice coordinates along the tube circumference and height, respectively. Additional details on these parameters are given in Ref. \cite{Didraga2002}.

For sheet geometries  the CAPs are applied along the brick width ($x$ direction) and transport along the $y$ direction, such that $w_n = y_n$. This choice ensures that both the number of transport channels and the spacing between absorbing potentials remain constant even in the presence of $slip$. For tubular aggregates, the CAPs are applied along the tube rotational axis  and transport along the tube height( z direction) such that,$w_n = z_n$. Channels in the rotational direction typically have low dimensionality (2–10), which makes CAP implementation along that direction impractical. A schematic of what this looks like for these geometries is shown in Fig \ref{fig:caps_schematic}.

The coordinates $w_{\min}$ and $w_{\max}$ denote the minimum and maximum positions along the chosen transport direction. The parameter $d_w$ defines the length of the left and right absorbing regions. In the $y$ direction, $y_{\min,n_2}$ and $y_{\max,n_2}$ correspond to the minimum and maximum coordinates of the $n_2^{\mathrm{th}}$ channel along $x$. This definition ensures that $d_y$ remains constant across all channels:

\begin{equation}
y_{\min,n_2} = \text{S} \cdot (n_2-1),
\end{equation}

\begin{equation}
y_{\max,n_2} = \text{L} \cdot (n_y - 1) + \text{slip} \cdot (n_2-1),
\end{equation}

\begin{equation}
d_y =L \cdot(\Omega-1),
\end{equation}

\begin{equation}
z_{\min} = 0,
\end{equation}

\begin{equation}
z_{\max} = \text{H} \cdot (n_z - 1),
\end{equation}

\begin{equation}
d_z = \text{H} \cdot (\Omega-1).
\end{equation}
 \( \Omega \)  is the unit length of the monomer of the left and right potentials.  A visual representation of CAPs on molecular aggregate systems for sheets and tubes is shown in Fig. \ref{fig:caps_schematic}(a).

\FloatBarrier
The energy-dependent transmission probability \cite{Seideman1992, CollepardoGuevara2004} is
\begin{equation}
T(E) = 4 \,\mathrm{Tr} \left[ \frac{1}{E - H + iV_I} \, V_{R} \, \frac{1}{E - H - iV_I} \, V_{L} \right]
= 4 \,\mathrm{Tr} \left[ G(E) \, V_{R} \, G^\dagger(E) \, V_{L} \right],
\label{eq:transmission}
\end{equation}
where $G(E)$ is the Green's function. To characterize transport, we define the thermally averaged transmission
\begin{equation}
\bar{T} = \frac{1}{Z} \sum_{i}^{n_g} T(\epsilon_i)\, e^{-\beta \epsilon_i}
= \frac{1}{Z} \int \sum_i \delta(\epsilon_i - E)\, e^{-\beta E}\, T(E)\, dE
= \frac{1}{Z} \int \rho(E)\, e^{-\beta E}\, T(E)\, dE,
\label{eq:thermal_transmission}
\end{equation}
with partition function
\begin{equation}
Z = \sum_i e^{-\beta \epsilon_i}
= \int \rho(E)\, e^{-\beta E}\, dE,
\label{eq:partition_function}
\end{equation}
where $\beta$ is the inverse temperature and $\rho(E)$ is the density of states. The quantity $\bar{T}$ measures the overall transmission of the aggregate.

\subsection{\textbf{Angle-Dependent 2D CAP Potential}}
To model two-dimensional exciton transport, we construct angle-dependent CAPs from the lattice coordinates of each molecular brick. The procedure is:

\begin{enumerate}
    \item Store the lattice coordinates and center them in the $x$--$y$ plane
    \[
    A(x_n, y_n) \rightarrow B = A - A_{\mathrm{mid}} .
    \]

    \item Convert the centered coordinates to cylindrical coordinates
    \[
    B \rightarrow C(R,\theta), \qquad
    R = \sqrt{x_n^2 + y_n^2}, \qquad
    \theta_n = \arctan\left(\frac{y_n}{x_n}\right).
    \]

    \item Select lattice points lying on a circle of radius $R_a$ within a specified tolerance.

    \item Construct the CAPs on the upper and lower arcs of the resulting circle of radius $R_a$.
\end{enumerate}

\FloatBarrier
\begin{figure}[htbp]
    \centering
\includegraphics[width=0.5\linewidth]{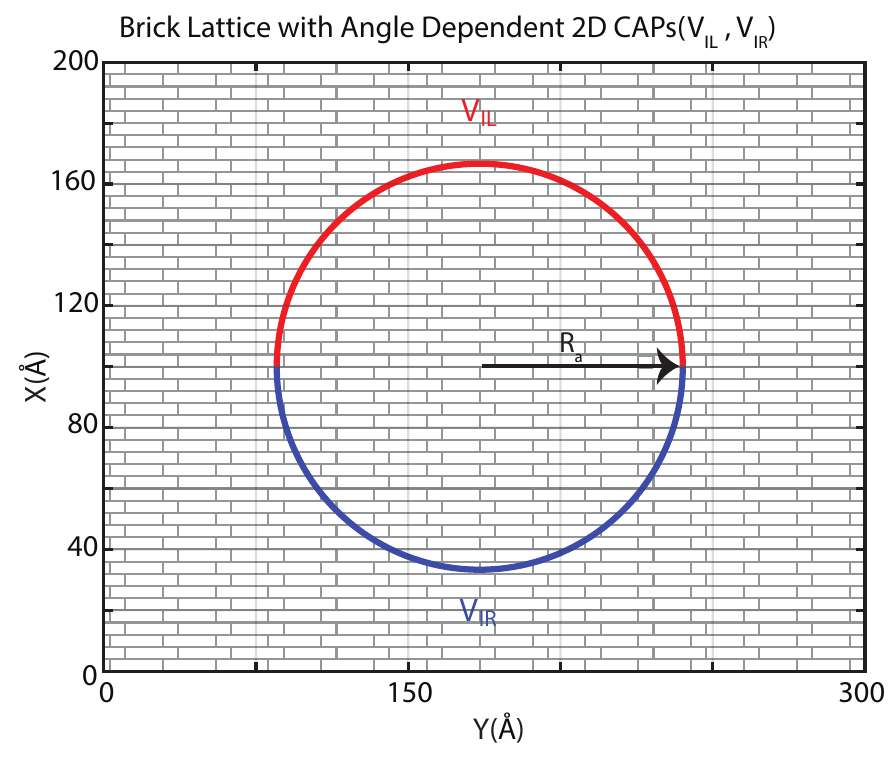}
    \caption{ Schematic of the 2D  circular absorbing potentials applied to a 2D sheet geometry. The circular boundary defines the absorbing region of radius $R_a$.}
    \label{fig:2d_caps_schematic}
\end{figure}
The angle-dependent CAP potentials are defined as
\begin{equation}
V_{IL}(\theta_n, \theta_0) =
\begin{cases}
V_{0I}\,\sin^{2}(\theta_n - \theta_0), & 0 \leq \theta_n < \pi, \\
0, & \text{otherwise}
\end{cases}
\label{eq:VIL}
\end{equation}
\begin{equation}
V_{IR}(\theta_n, \theta_0) =
\begin{cases}
V_{0I}\,\sin^{2}(\theta_n - \theta_0), & \pi \leq \theta_n < 2\pi, \\
0, & \text{otherwise}
\end{cases}
\label{eq:VIR}
\end{equation}
Here $\theta_0$ is a shift angle that controls coherent transport in the system. A schematic of the CAPs is shown in Fig. \ref{fig:2d_caps_schematic}. With $\theta_0$, we can tune which Bloch states contribute to the transmission. 
\subsection{Reducing Computational Cost} 
Experimentally relevant aggregates can extend hundreds of nanometers, and their properties converge only for sufficiently large systems. Conventional transmission calculations scale as $\mathcal{O}(N^3)$ due to matrix inversion and multiplication. To reduce this cost, we employ stochastic sampling with Chebyshev polynomial expansion to evaluate the Green's function \cite{CollepardoGuevara2004,Mandelshtam1995,deCastro2023}. The Hamiltonian is applied efficiently using 2D convolution, as previously demonstrated for molecular aggregates \cite{Bradbury2020,Bradbury2024}. This approach reduces the computational scaling from $\mathcal{O}(N^3)$ to $\mathcal{O}(N\log N)$.

The transmission probability can be written as
\begin{equation}
T(E) = 4\,\mathrm{Tr}\!\left[A^\dagger(E)A(E)\right],
\end{equation}
\begin{equation}
A(E)= \hat{V}_R^{1/2} G(E) \hat{V}_L^{1/2}.
\end{equation}
Using stochastic trace evaluation and Chebyshev expansion, $T(E)$ becomes
\begin{equation}
T(E) \approx \frac{4}{N_{\text{stoc}}}\sum_{i=1}^{N_{\text{stoc}}}
\langle \chi_i | A'^\dagger A' | \chi_i \rangle,
\end{equation}
\begin{equation}
A'(E)= \hat{V}_R^{1/2}
\left(\sum_{n=0}^{N_{\text{chb}}} a_n(E)\,\hat{Q}_n(\hat{H}',\hat{V}_I)\right)
\hat{V}_L^{1/2}.
\end{equation}
The stochastic vectors $\chi_i$ have elements randomly chosen as $\pm1$. The matrices $\hat{V}_R^{1/2}$ and $\hat{V}_L^{1/2}$ are the square roots of the diagonal elements of $\hat{V}_R$ and $\hat{V}_L$. The coefficients $a_n(E)$ are the Chebyshev expansion coefficients, and $\hat{Q}_n(\hat{H},\hat{V}_I)$ are the associated Chebyshev polynomials.

The operator
\[
\sum_{n=0}^{N_{\text{chb}}} a_n(E)\hat{Q}_n(\hat{H}',\hat{V}_I)
\]
represents the action of the Green's function on an arbitrary vector. The coefficients are

\begin{equation}
a_n= \frac{i(2-\delta_{n,0})}{\Delta H\sin\phi}e^{-in\phi},
\end{equation}

\[
\phi = \arccos\left(\frac{E-\bar{h}}{\Delta H}\right).
\]

Here $\bar{h}$ and $\Delta H$ scale the Hamiltonian so that the eigenvalues of $H'$ lie in $[-1,1]$,

\begin{equation}
\bar{h}=\frac{\epsilon_{\max}+\epsilon_{\min}}{2},
\end{equation}

\begin{equation}
\Delta H=\frac{\epsilon_{\max}-\epsilon_{\min}}{2},
\end{equation}

\begin{equation}
H'=\frac{H-\bar{h}}{\Delta H}.
\end{equation}
The quantities $\epsilon_{\max}$ and $\epsilon_{\min}$ denote the any upper and lower bounds ,repetitively, of $H$. The Chebyshev recursion is given by

\begin{equation}
\chi^{\gamma}_0=\hat{Q}_0(\hat{H},\hat{V}_I)\chi^{ext}=\chi^{ext},
\end{equation}

\begin{equation}
\chi^{\gamma}_1=\hat{Q}_1(\hat{H},\hat{V}_I)\chi^{ext}
= e^{-\gamma}\hat{H}'\chi^{ext} \quad
\gamma=\frac{\hat{V}_I}{\Delta H}.
\end{equation}

and for $n>1$,

\begin{equation}
\chi^{\gamma}_n=e^{-\gamma}\left(2\hat{H}'\chi^{\gamma}_{n-1}-e^{-\gamma}\chi^{\gamma}_{n-2}\right).
\end{equation}

\section{Results/Discussion}
\subsection{Simulation Specifications for CAPs }
The parameters for the angle-dependent CAPs and the one-dimensional (1D) CAPs are listed in Tables \ref{table:1D_para} and \ref{table:2D_para}.

\begin{table}[h]
\centering
\begin{tabular}{|l|c|c|c|c|}
\hline
$V_{OI}\,(\mathrm{cm}^{-1})$ 
& $\Omega$ (Monomer Units) 
& $N_{\text{stoc}}$ 
& $N_{\text{chb}}$ 
& Temperature (K) \\
\hline
2130 
& $\tfrac{n_z}{4}$ (tubes), $\tfrac{n_y}{4}$ (sheets) 
& 60 
& 2000 
& 300 \\
\hline
\end{tabular}
\caption{Parameters used for the 1D CAP simulations in tube and sheet geometries.}
\label{table:1D_para}
\end{table}

\begin{table}[h]
\centering
\begin{tabular}{|l|c|c|c|c|}
\hline
$V_{OI}\,(\mathrm{cm}^{-1})$ 
& Radius of CAP (monomer units) 
& $N_{\text{stoc}}$ 
& $N_{\text{chb}}$ 
& Temperature (K) \\
\hline
2130 
& 20 
& 1200 
& 2000 
& 300 \\
\hline
\end{tabular}
\caption{Parameters used for the 2D angle-dependent CAP simulations in sheet aggregates.}
\label{table:2D_para}
\end{table}

\subsection{Predicting Delocalization Behavior with CAPs}

Before using CAPs to measure exciton transport, we first tested whether they reproduce the delocalization behavior of known disorder effects. To do so, we compare the average transmission and the participation ratio (PR) in the presence of energetic disorder. PR quantifies exciton delocalization and is closely related to transport efficiency \cite{Sneyd2021,Giannini2022}.

\begin{figure}[htbp]
    \centering
    \includegraphics[width=0.5\linewidth]{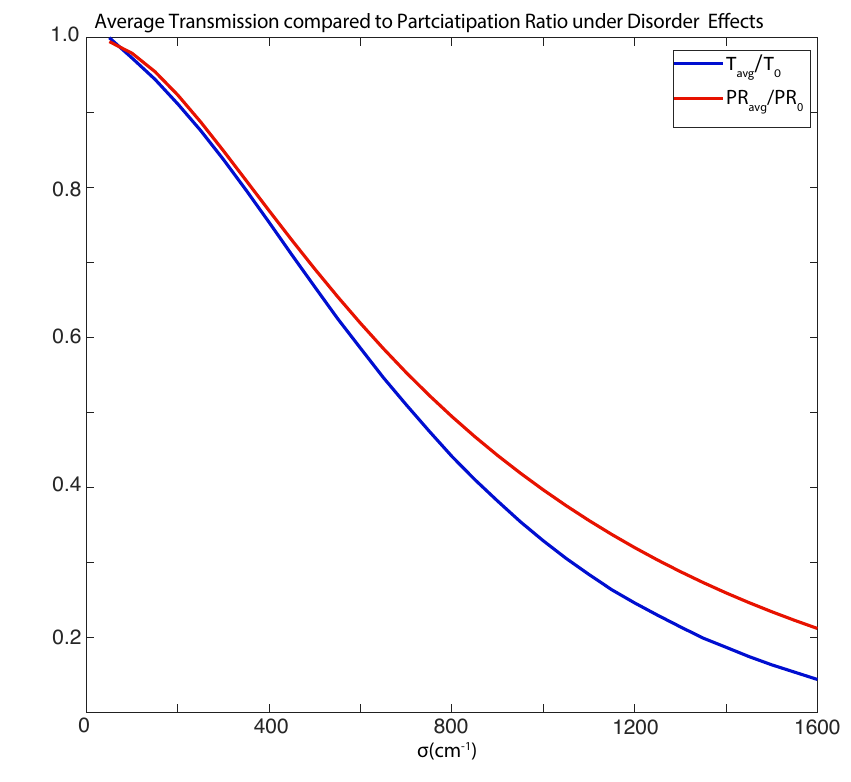}
    \caption{Average participation ratio ($PR_{\mathrm{avg}}$) and average transmission ($T_{\mathrm{avg}}=\bar{T}$) as functions of disorder strength ($\sigma$). Both quantities are normalized to their respective values at zero disorder, $PR_0=PR_{avg}(\sigma=0)$ and $T_0=T_{avg}(\sigma=0)$. Results are shown for a 2D sheet system of size $20 \times 100$.}
    \label{fig:ipr_transmission_disorder}
\end{figure}

Fig. \ref{fig:ipr_transmission_disorder} compares the PR with the average transmission computed using the 1D CAP potential as the energetic disorder strength ($\sigma$) increases. The transmission reproduces both the trend and overall shape of the PR. Thus, the average transmission predicts exciton delocalization and can serve as an alternative to PR/IPR(Inverse Participation ratio), with the additional advantage of providing directional information about transport.

\FloatBarrier
\subsection{CAPs: Vacancy and Size Effects }
\begin{figure}[htbp]
    \centering
    \includegraphics[width=0.75\linewidth]{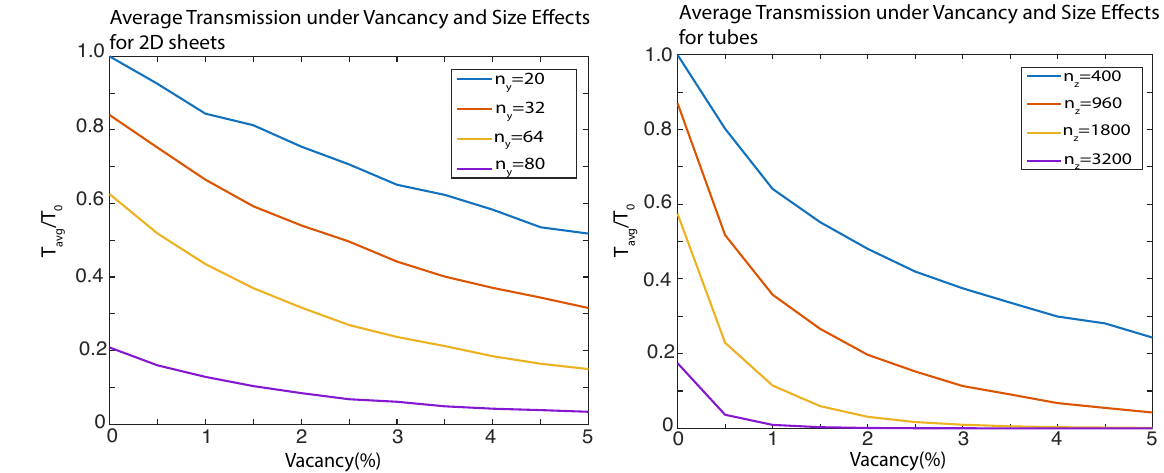}   
    \caption{ (a) Plot of Average Transmission for 2D sheet aggregate with varying $n_y$ and vacancy percentage. $n_{x}$ is growing alongside $n_x$, keeping the ratio of $n_x:n_y$ at 5:1, meaning $n_x=5n_y$ (b) Plot of Average Transmission for tubular aggregate with varying $n_z$ and vacancy percentage. $n_{rot}$ is constant at 5.}
    \label{fig:vac_size}
\end{figure}

We examine the effect of vacancy and system size on transmission. Fig \ref{fig:vac_size} shows plots of the average transmission as a function of size and vacancy for both sheets and tubes, respectively. In both plots, the longer the path traversed by excitons in the presence of vacancies, the more the transmission decreases. This indicates that transmission is hindered more by the absolute number of sites removed than by the ratio of vacant sites to the number of monomers in the aggregate. The figure also shows that sheets are more resistant to vacancy effects than tubes. These results mirror the findings on diagonal disorder and dimensionality reported in \cite{Chuang2016}, where 2D structures were shown from IPR calculations to be resistant to diagonal disorder than pseudo-1D tubes. 
\FloatBarrier

\subsection{Transmission Calculations Using 2D CAPs}
\begin{figure}[htbp]
    \centering
    \includegraphics[width=0.75\linewidth]{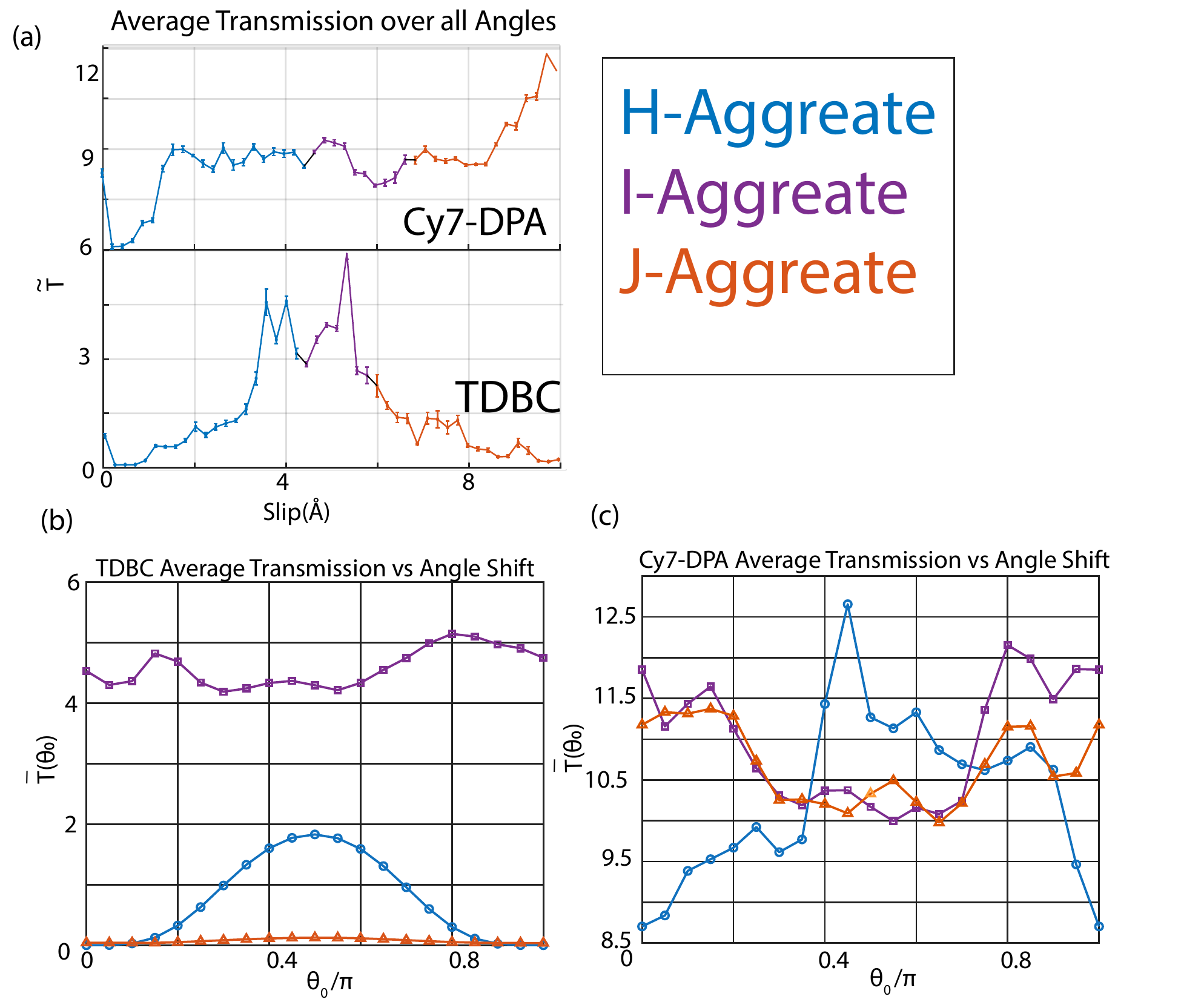}
    \caption{Transmission calculated using angle-dependent CAPs with Aggregate Classification of H,J, I aggregate. (a) Angle-averaged transmission, $\tilde{T}$. 
    (b) Average transmission as a function of shift angle, $\bar{T}(\theta_0)$, for TDBC. 
    (c) Average transmission as a function of shift angle, $\bar{T}(\theta_0)$, for Cy7--DPA. 
    Panels (b) and (c) show the angular dependence across different aggregate regimes (H-, I-, and J-aggregates), classified according to the slip parameter.}
    \label{fig:angle_dependent_transmission}
\end{figure}

Next, we computed the angle-averaged transmission using the angle-dependent CAPs defined in Eqs. (\ref{eq:VIL}) and (\ref{eq:VIR}). The shift angle $\theta_0$ and the slip parameter are varied across the H-, J-, and I-aggregate regimes for Cy7--DPA and TDBC. Screening this angle in different regimes allows us to see which bloch waves contribute the most to the transport. For each slip value, the transmission is averaged over all sampled angles according to Eq.~(\ref{eq:angle_averaged_transmission}).

\begin{equation}
\tilde{T} = \frac{1}{\pi} \int_{0}^{\pi} \bar{T}(\theta_0)\, d\theta_0. 
\label{eq:angle_averaged_transmission}
\end{equation}
This measure quantifies how the shift angle influences transport for a given slip, identifying regimes in which angular tuning either enhances, suppresses, or leaves the transmission largely unchanged. Fig. \ref{fig:angle_dependent_transmission}(a) shows $\tilde{T}$ for both aggregates as a function of slip. Two transmission regimes emerge: one with suppressed transmission and another with enhanced transmission. We denote these regimes as \textit{I.S.-aggregates} (insulating-type aggregates) and \textit{S-aggregates} (semiconducting-type aggregates), respectively. The classification is aggregate-specific, and the corresponding slip ranges are listed in Table \ref{tab:slip_regimes}.

\FloatBarrier
\begin{table}[htbp]
\centering
\begin{tabular}{|c|c|c|}
\hline
Aggregate & S-Aggregate Slip Regime ($\mathrm{\AA}$) & I.S.-Aggregate Slip Regime ($\mathrm{\AA}$) \\
\hline
Cy7-DPA & 2.0--10 & 0.0--2.0 \\
\hline
TDBC     & 3.6--4.4 & 0--3.6, 4.4--10 \\
\hline
\end{tabular}
\caption{Slip parameter ranges defining the S- and I.S.-aggregate regimes for Cy7--DPA and TDBC.}
\label{tab:slip_regimes}
\end{table}
\FloatBarrier
The following classification between H, J, and I for TDBC and Cy7-DPA is based on the slip value for each aggregate, as referenced in \cite{Deshmukh2022}. For TDBC, the I-aggregate regime is entirely within the S-aggregate classification. For Cy7–DPA, all of the  I- and J-aggregate regimes belong to the S-aggregate classification.

Fig. \ref{fig:angle_dependent_transmission}b and c show the averaged transmission dependent on the angle, $\bar{T}(\theta_0)$, for both TDBC and Cy7--DPA. The figure shows maximum transmission in the H-aggregate regime at $\theta_0=\frac{\pi}{2}$. Moving away from this angle dramatically reduces the transmission, unlike in other aggregate types. In contrast, the I-aggregate regime exhibits the smallest variation in transmission across shift angles.

To rationalize the trends observed above, we analyze the contributions of individual $k$-states within the S- and I.S.-aggregate regimes as a function of $\theta_0$. The contribution from each $k$-state to the transmission is quantified by
\begin{equation}
\bar{T}(k_x, k_y, \theta_0)= 
\frac{1}{Z}\int \rho(E)\, e^{-\beta E}\, 
T(k_x, k_y, E, \theta_0)\, dE,
\label{eq:k_resolved_Tbar}
\end{equation}

\begin{equation}
T(k_x, k_y, E, \theta_0)
= 4 \, \langle k_x, k_y \vert 
A(E, \theta_0)\, A^{\dagger}(E, \theta_0) 
\vert k_x, k_y \rangle ,
\label{eq:k_resolved_T}
\end{equation}
with the Bloch wave states $\lvert k_x, k_y \rangle$ defined as
\begin{equation}
\langle j_x, j_y \mid k_x, k_y \rangle
=
\frac{1}{\sqrt{N}}
\exp\!\left( i 2\pi \frac{k_x j_x}{n_x} \right)
\exp\!\left( i 2\pi \frac{k_y j_y}{n_y} \right),
\label{eq:bloch_state_2d}
\end{equation}
where $k_x = 0,1,2,\dots,(n_x-1)$ and 
$k_y = 0,1,2,\dots,(n_y-1)$.

\begin{figure}[h]
    \centering
    \includegraphics[width=0.75\linewidth]{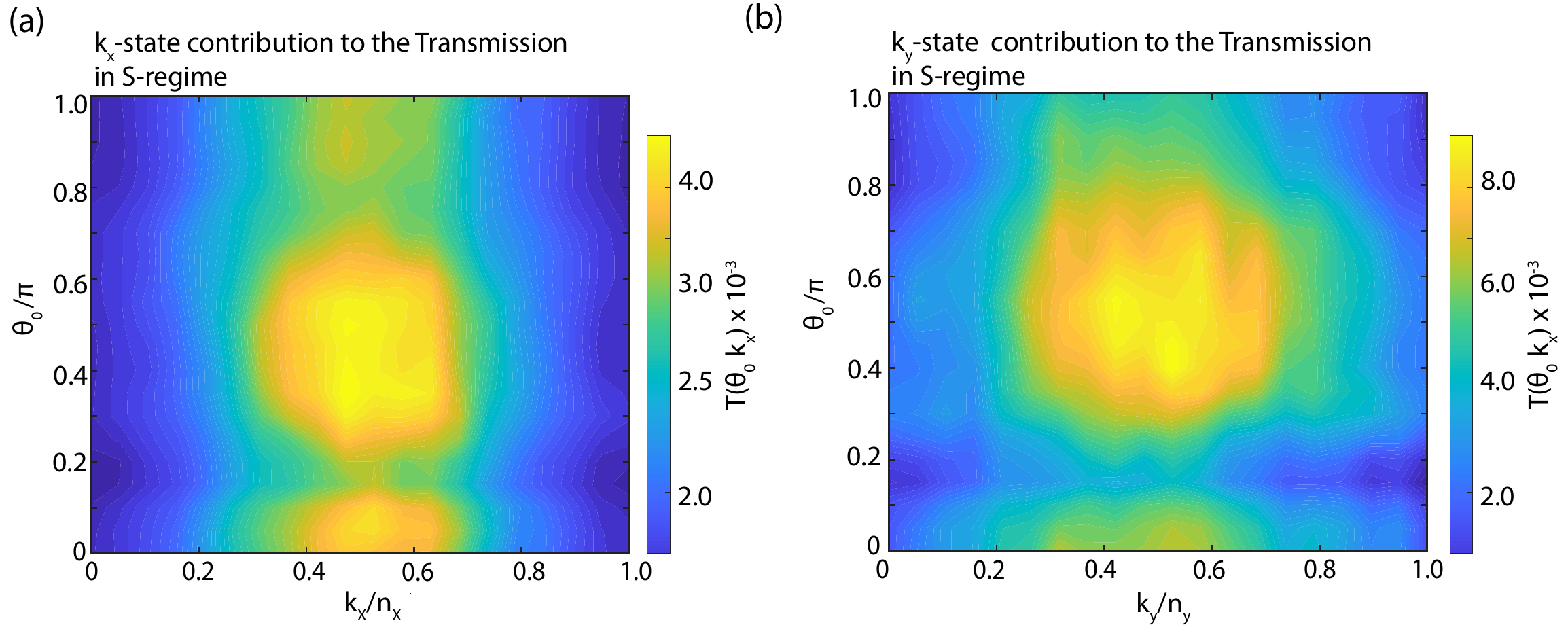}
    
    \vspace{0.4cm}
    
    \includegraphics[width=0.75\linewidth]{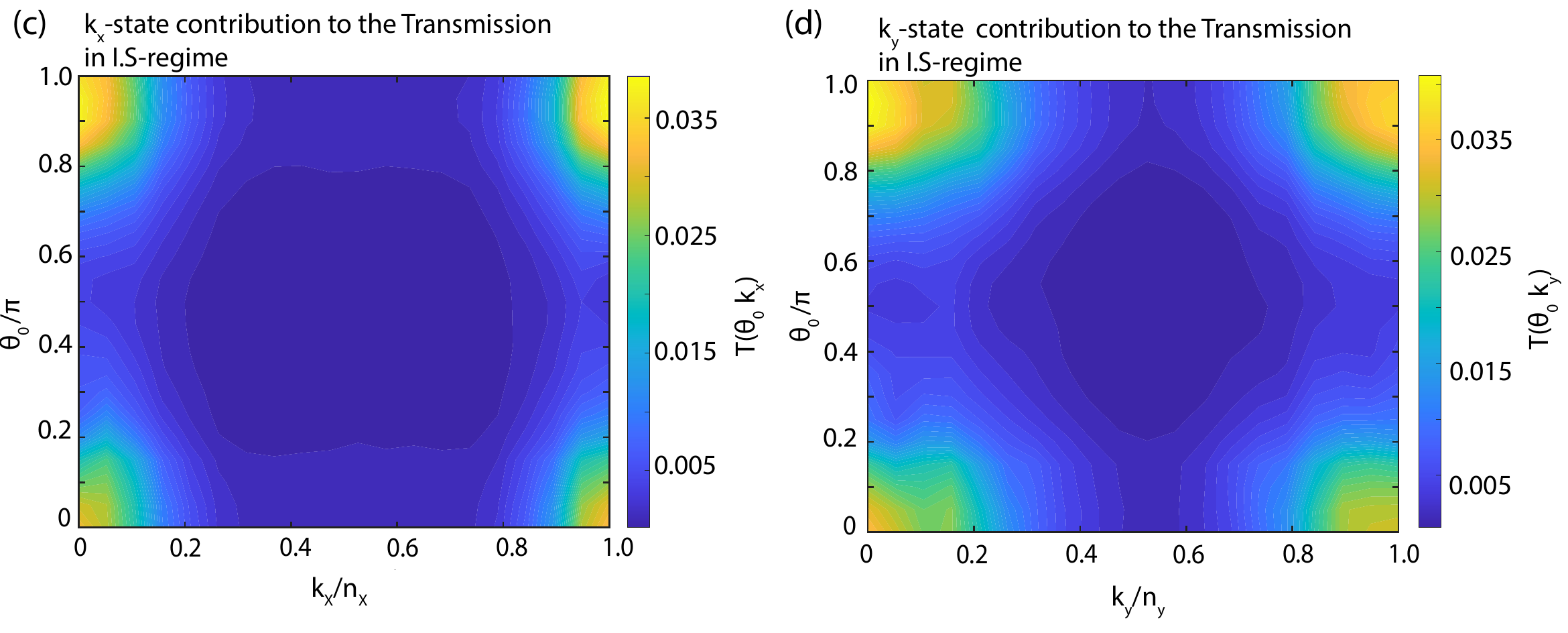}
    
    \caption{   k-state contribution of transmission for contour plots of TDBC under varying $\theta_0$ in S and I.S regimes.  
    Top: S-aggregate regime (slip = 4 \AA). (a) $k_x$ (b) $k_y$
    Bottom: I.S-aggregate regime (slip = 9 \AA). (c) $k_x$ (d) $k_y$
    In each case: For $k_x$ contribution,  $k_y=0$, 
    (b) For $k_y$ contribution, $k_x=0$.}
    
    \label{fig:TDBC_k_combined}
\end{figure}

\begin{figure}[h]
    \centering
    
    \includegraphics[width=0.75\linewidth]{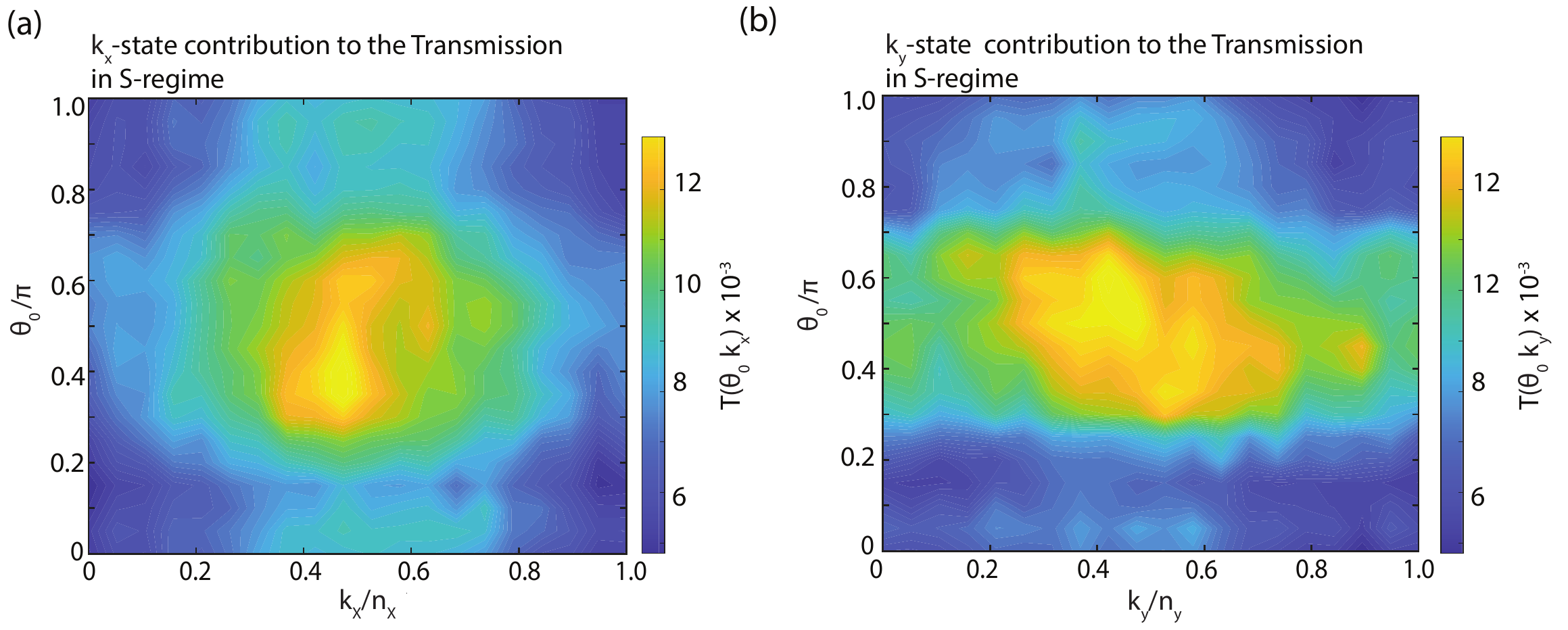}
    
    \vspace{0.4cm}
    
    \includegraphics[width=0.75\linewidth]{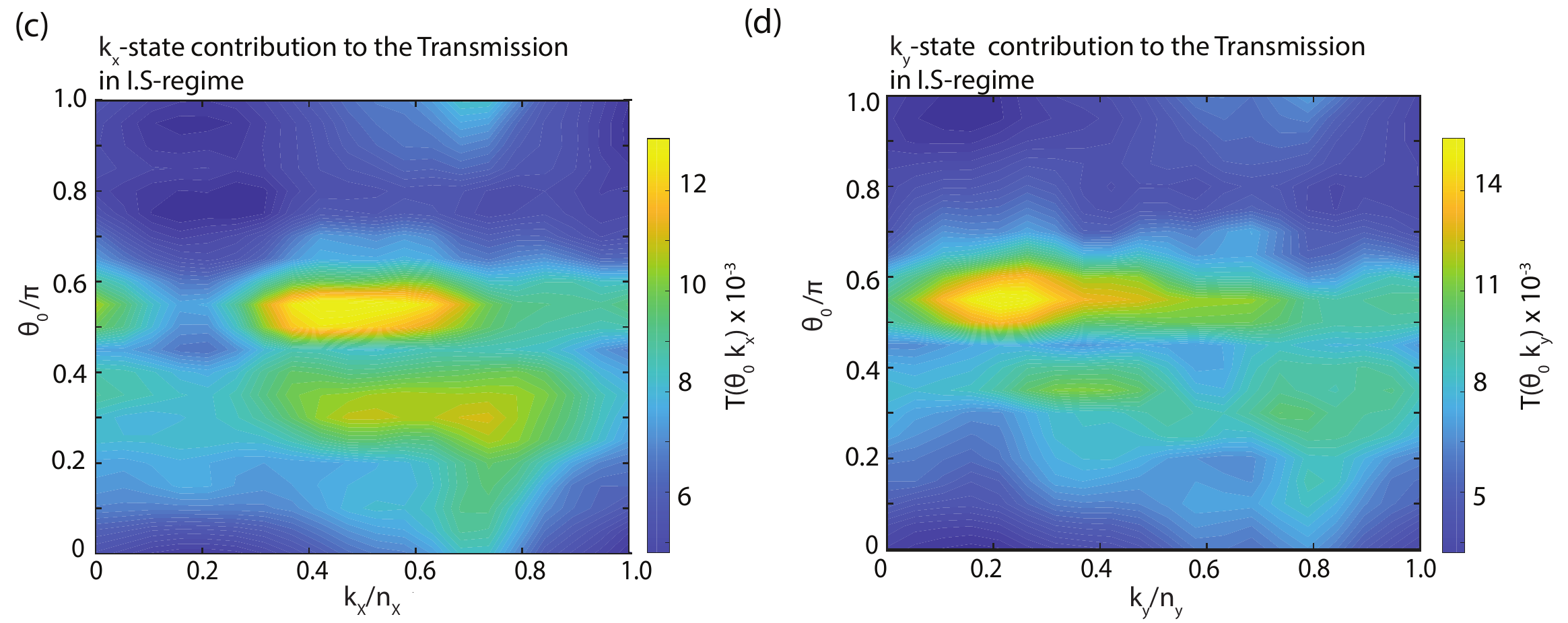}
    
    \caption{
    k-state contribution of transmission for contour plots of Cy7-DPA under varying $\theta_0$ in S and I.S regimes. 
    Top: S-aggregate regime (slip = 4 \AA). (a) $k_x$ (b) $k_y$
    Bottom: I.S-aggregate regime (slip = 1 \AA). (c) $k_x$ (d) $k_y$
    In each case: For $k_x$ contribution,  $k_y=0$, 
    (b) For $k_y$ contribution, $k_x=0$.}
    
    \label{fig:Cy7_k_combined}
\end{figure}

Fig. \ref{fig:TDBC_k_combined} and \ref{fig:Cy7_k_combined} show the contributions of individual $k$-states calculated using Eq.~(\ref{eq:k_resolved_Tbar}). As illustrated in these figures, the control parameter $\theta_0$ selects the subset of $k$-states that dominate the transmission.  For the S-aggregate regime in both TDBC and Cy7--DPA, the transmission is maximized by states near $(k_x, k_y) = \left(\frac{n_x - 1}{2}, \frac{n_y - 1}{2}\right)$ when $\theta_0 = \pi/2$. In contrast, the dominant $k$-states in the I.S.-aggregate regime differ between the two systems. For TDBC, the largest contributions arise from states near $(k_x, k_y) = (0, n_y - 1)$ and $(n_x - 1, 0)$ when $\theta_0 = 0$ and $\pi$. For Cy7-DPA, the maximum transmission in the I.S.-aggregate regime also occurs near $(k_x, k_y) = \left(\frac{n_x - 1}{2}, \frac{n_y - 1}{2}\right)$ at $\theta_0 = \pi/2$, but with a substantially smaller number of contributing states compared to the S-aggregate regime. In general, transmission is maximized when contributing $k$-states are located near the center of the Brillouin zone and is further enhanced when $\theta_0 = \pi/2$.

\FloatBarrier
\section{Conclusion}
This work demonstrates the capability of complex absorbing potentials (CAPs) to probe exciton dynamics in molecular aggregate sheets and tubes, and to compare the spectral characteristics associated with different types of disorder. Stochastic methods enable the screening of larger systems, bringing the results closer to experimentally relevant length scales. Using these techniques, the effects of vacancies and energetic disorder on exciton transport in two-dimensional aggregate sheets are successfully modeled. The results indicate that reduced transmission correlates with diminished transport, allowing us to predict how vacancies and disorder impede exciton motion. We also identify two distinct aggregate regimes based on transmission behavior and, consequently, coherent transport properties. The angle-dependent CAP framework provides a systematic approach for linking molecular packing, aggregate classification, and transport efficiency in two-dimensional molecular aggregates.

In future work, these methods will be extended to additional systems and combined with complementary approaches, including cavity-coupled systems and biexcitonic systems. Using  CAPs in time-evolution studies for molecular aggregates has the advantage of suppressing backscattering at  boundaries, enabling more accurate predictions of time-dependent quantities such as the diffusion constant, localization length, and average travel time.
\bibliographystyle{achemso}
\bibliography{ref}

\end{document}